\documentclass[preprint,aps,prd,floatfix]{revtex4-2}

\usepackage{graphicx}
\usepackage{amsfonts}
\usepackage{amsmath}
\usepackage{rotating}
\usepackage{amssymb,amsthm}
\usepackage{soul}

\usepackage{xcolor}
\usepackage[normalem]{ulem}

\theoremstyle{definition}

\theoremstyle{remark}

\newcommand{\mbold}[1]{\mbox{\boldmath{\ensuremath{#1}}}}

\def\be{\begin{equation}}
\def\ee{\end{equation}}
\def\beq{\begin{eqnarray}}
\def\eeq{\end{eqnarray}}

\def \bell {\mbox{{\mbold\ell}}}
\def \bk {\mbox{{\bf k}}}

\def \bm {\mbox{{\bf m}}}
\def \bh {\mbox{{\bf h}}}
\def \bomega {\mbox{{\mbold \omega}}}

\begin{document}

\title{Teleparallel Geometry with a Single Affine Symmetry}

\author {A. A. Coley}
\email{aac@mathstat.dal.ca}
\affiliation{Department of Mathematics and Statistics, Dalhousie University, Halifax, Nova Scotia, Canada, B3H 3J5}

\author{R. J. {van den Hoogen}}
\email{rvandenh@stfx.ca}
\affiliation{Department of Mathematics and Statistics, St. Francis Xavier University, Antigonish, Nova Scotia, Canada, B2G 2W5}



\begin{abstract}
In teleparallel geometries, symmetries are represented by affine frame symmetries which constrain both
the (co)frame basis and the spin-connection (which are the primary geometric objects).
In this paper we shall study teleparallel geometries with a single affine symmetry,
utilizing the locally Lorentz covariant approach and adopting a complex null gauge.  
We first introduce an algorithm to study geometries with an affine frame symmetry, which consists of
choosing coordinates adapted to the symmetry, constructing a canonical frame, and solving the equations describing the symmetry.
All of the constraints on the geometry are determined in the case of a single affine symmetry, but there are additional constraints
arising from the field equations for a given theory of teleparallel gravity.
In particular, we find that in $f(T)$ teleparallel gravity there will be severe constraints on the geometry arising from the antisymmetric part of the field equations.
\end{abstract}

\maketitle

\tableofcontents

\newpage

\section{Introduction}

There is continued interest in alternative theories to Einstein’s General Theory of
Relativity (GR), partly related to issues of dark energy and dark matter in
cosmology \cite{Nojiri_Odintsov2006,Capozziello_DeLaurentis_2011}. In one class of gravity theories the dynamics of the gravitational field are assumed to be encoded in the torsion of a spacetime with zero curvature \cite{Krssak2015,Bahamonde_Boehmer_Wright2015}. In particular, in so-called
teleparallel theories of gravity (or teleparallel gravity in brief)  
the non-metricity is also assumed to vanish, in which case there
is a subclass of teleparallel gravity (TEGR) that is dynamically equivalent to GR.
The Lagrangian for TEGR theory is based on a scalar,
$T$, constructed from the torsion tensor. Since the Lagrangian differs to that of GR by a total
derivative, the field equations are formally equivalent. We shall generally refer to
solutions of the field equations to teleparallel gravity theories as {\em{teleparallel geometries}}.

A topical generalization of  TEGR is so-called
$f(T)$ gravity \cite{Ferraro:2006jd,Ferraro_Fiorini2011}.
If the teleparallel geometry is defined in a gauge invariant manner,
then the most general spin-connection with zero curvature is the purely inertial
spin-connection, which vanishes in the special class of “proper" frames where
all inertial effects are absent; the spin-connection is non-zero in all other frames  \cite{Obukhov_rubilar2006,Lucas_Obukhov_Pereira2009,Aldrovandi_Pereira2013,Krssak:2018ywd}. The advantage
of the covariant approach is that by using the purely inertial connection, the resulting teleparallel gravity theory embodied by the Lorentz covariant field equations is locally
Lorentz invariant \cite{Krssak:2018ywd}, and it is then admissible to use an arbitrary tetrad in an arbitrary coordinate system
with the corresponding spin-connection to produce equivalent field equations to those in the
proper frame \cite{Krssak_Saridakis2015,Krssak_Pereira2015}.

In theories of this type, the metric tensor is replaced by the frame basis as the fundamental geometric object.  
To determine solutions of
the corresponding field equations for a particular teleparallel gravity theory, 
a coordinate system,
$x^\mu$, a coframe basis, $\bh^a$, and a spin-connection, $\omega^a_{~bc}$  (or, alternatively,  a proper coframe basis in
which the spin-connection is trivial) must be chosen. It is possible that two seemingly distinct choices of the
coordinates, coframe basis and spin-connection which satisfy the teleparallel field equations
are, in fact, the same solution, but this fact can be hidden by these choices. It is necessary to
uniquely characterize a solution in an invariant manner.
To determine the equivalence of two teleparallel geometries, a
modification of the Cartan-Karlhede algorithm adapted to Riemann-Cartan geometries has been introduced \cite{CMT}. This necessitates an invariant frame, where the frame basis elements are completely fixed. Other frames can, of course, be utilized, an obvious example being the proper
frame in which the spin-connection is trivial, $\omega^a_{~bc}=0$
(however, this frame is not an invariant frame).

It has been also shown that inequivalent solutions to the teleparallel field equations can give rise to
equivalent metrics. This implies that for a given frame distinct
teleparallel geometries can be produced  by a choice of different spin-connections which are not related by a Lorentz frame transformation. This motivates the investigation
of spin-connections which generate different torsion geometries with the same metric. In particular, two teleparallel geometries can be
shown to be inequivalent by comparing their group of symmetries (for example,
isometries, which are symmetries of the metric and are necessarily coordinate-independent) \cite{Hecht:1992xn}.
We notice that the symmetry group of the solutions of a teleparallel gravity theory is potentially
smaller than their metric-based analogues in GR \cite{CMT}.

\newpage


\subsection{Symmetry}

The role of an isometry, which is a diffeomorphism from the space into itself which preserves the metric,
in covariant teleparallel gravity theories is not as clear as in metric
based theories. In a teleparallel geometry, the tetrad (or (co)frame) and the zero curvature spin-connection,
replace the metric as the principal objects of study since the frame and spin-connection are
used in the calculation of the torsion tensor and the field equations. The metric is now a
derived geometrical object constructed from symmetric products of the frame elements. The symmetries of a particular teleparallel geometry may not coincide with the set of
isometries. Indeed, an affine frame symmetry is an isometry (i.e., a Killing vector field, ${\bf X}$, where
$\mathcal{L}_{{\bf X}} g = {\bf 0}$),
but not all isometries are necessarily affine frame symmetries.

An affine frame symmetry is a diffeomorphism  on $\mathcal{F}(M)$, the frame bundle of the manifold $M$ \cite{olver1995}, such that there exists a frame in which the
vector field, ${\bf X}$, satisfies \cite{fon1992}:
\begin{equation}
\mathcal{L}_{{\bf X}} \bh_a = 0 \text{ and } \mathcal{L}_{{\bf X}} \bomega^a_{~b} = 0 \label{Intro:FS2} 
\end{equation}
(which thus leaves the frame basis and the associated spin-connection invariant), where $\omega^a_{~bc}$ denotes the spin-connection corresponding to the 
geometrically preferred invariant frame, $\bh_a$. 
This characterization is \emph{a frame-dependent} analogue of the definition of a symmetry introduced by \cite{HJKP2018,Pfeifer22} (see also \cite{Hohmann:2015pva}).
If the solution admits a non-trivial isotropy group, we are able to \emph{prolong} the manifold \cite{olver1995} to produce a larger manifold and determine an invariant frame \cite{MCH}.

In this paper we wish to study spacetimes with a single affine frame symmetry, ${\bf X}$.  We shall study spacetimes with multiple affine frame symmetries having no isotropies \cite{HMC} (e.g., simply transitive spatially homogeneous spacetimes; see also \cite{Hohmann:2020zre}) and spacetimes with isotropies \cite{MCH} (e.g., spherically symmetric spacetimes; see also\cite{Golovnov2}) elsewhere.

We shall follow the notation of \cite{CMT}. 
The coordinate indices are denoted by $\mu, \nu, \ldots$, the tangent space indices are denoted by $a,b,\ldots$, and the spacetime coordinates are denoted by $x^\mu$. The frame fields are $\bh_a$ and the dual coframe one-forms are $\bh^a$, and the corresponding vielbein components are denoted by $h_a^{~\mu}$ and $h^a_{~\mu}$, respectively. For a given anholonomic coframe, $\bh^a$, the structure coefficients are denoted by  $C^c_{~ab}$. The spacetime metric is $g_{\mu \nu}$, and the Minkowski tangent space metric is $\eta_{ab}$.
We will write $\Lambda^a_{~b}(x^\mu)$
to denote a local Lorentz transformation leaving $\eta_{ab}$ invariant. The spin-connection one-form $\bomega^a_{~b}$, is designated by $\bomega^a_{~b} = \omega^a_{~bc} \bh^c$. We will work with the torsion two-form, ${\bf T}^a = \frac12 T^a_{~bc} \bh^b \wedge \bh^c$.

\subsection{Gauge choices and Lorentz transformations}

We will assume that the field equations are covariant under local $GL(4,\mathbb{R})$ gauge transformations. We take advantage of this gauge freedom of local linear transformations in the tangent space to judiciously simplify aspects of the calculations. We shall adopt the {\it complex null gauge} with complex null coframe $ \{ \bh {^a} \} = \{ {\bf k}, \bell, \bm, \bar{\bm} \}$ \cite{classc}, in which
\beq
g_{ab} = \eta_{ab}:=\left[ \begin{array}{cccc} 0 & -1 & 0 & 0 \\ -1 & 0 & 0 & 0 \\ 0 & 0 & 0 & 1 \\ 0 & 0 & 1 & 0 \end{array}\right].
\label{gauge} \eeq

\noindent
[The alternative \emph{Orthonormal gauge} is a gauge choice that diagonalizes the tangent space metric $g_{ab}= \eta_{ab} = \mbox{Diag}[-1,1,1,1]$.]

There remains a $O(1,3)$ subgroup of $GL(4,\mathbb{R})$ of residual gauge transformations  of the coframe that leaves the metric invariant.  
We further restrict this subgroup to be the proper orthochronous Lorentz subgroup, $SO(1,3)$ of $O(1,3)$,
in order to preserve the orientation of space and the direction of time. We shall denote these Lorentz transformations of the coframe as: $$ \bh^a \to   {\bh'}^a=\Lambda^a_{~b} {\bh}^b. $$

The resulting field equations must transform homogeneously under this remaining gauge freedom, in this case $SO(1,3)$ Lorentz transformations. The Lorentz frame transformations within a complex null gauge are  \cite{kramer}:
\begin{itemize}
\item A null rotation about $\bell$: $\bell^{\prime} = \bell, {\bf k}^{\prime} = {\bf k} + \bar{E} {\bf m} + E{\bf \bar{m}} +E\bar{E} \bell,  {\bf m}^{\prime} = {\bf m} + E \bell,$ parameterized by the complex function $E(x^\mu)$.
\item A null rotation about ${\bf k}$: The effect of which can be determined similarly as above in terms of a complex function $B(x^\mu)$.
\item A boost and a spin parameterized by real functions $A(x^\mu)$ and $\theta(x^\mu)$.
\end{itemize}

The matrix representations of the corresponding Lorentz transformations acting on the coframe $ \{ \bh {^a} \} = \{ {\bf k}, \bell, \bm, \bar{\bm} \}$ are explicitly given by:
\begin{equation}
{}^l\Lambda^b_{~c}=\begin{bmatrix}
1 & E\bar{E} & \bar{E}  & E  \\
0 & 1 & 0 & 0   \\
0 & E & 1 & 0  \\
0 & \bar{E} & 0 & 1 \\
\end{bmatrix}
; \qquad
({}^l\Lambda^{-1})^{b}_{~c}=\begin{bmatrix}
1 & E\bar{E} &-\bar{ E} & -E  \\
0 & 1 & 0 & 0   \\
0 & -E & 1 & 0  \\
0 & -\bar{E} & 0 & 1 \\
\end{bmatrix},
\end{equation}
\begin{equation}
{}^k\Lambda^b_{~c}=\begin{bmatrix}
1 & 0 & 0 & 0  \\
B\bar{B} & 1 & \bar{B} & B   \\
B & 0 & 1 & 0  \\
\bar{B} & 0 & 0 & 1 \\
\end{bmatrix}
;\qquad
({}^k\Lambda^{-1})^{b}_{~c}=\begin{bmatrix}
1 & 0 & 0 & 0  \\
B\bar{B} & 1 & -\bar{B} & -B   \\
-B & 0 & 1 & 0  \\
-\bar{B} & 0 & 0 & 1 \\
\end{bmatrix},
\end{equation}
and
\begin{equation}
{}^A\Lambda^c_{~d}\,{}^{\theta}\Lambda^d_{~e} = \begin{bmatrix}
A & 0 & 0 &0  \\
0 & A^{-1} & 0 & 0   \\
0 & 0 & e^{i\theta} & 0  \\
0 & 0 &  & e^{-i\theta} \\
\end{bmatrix}
; \qquad ({}^{\theta}\Lambda^{-1}\,)^{c}_{~d}  ({}^A\Lambda^{-1})^{d}_{~e} =
 \begin{bmatrix}
A^{-1} & 0 & 0 & 0  \\
0 & A & 0 & 0   \\
0 & 0 & e^{-i\theta} & 0  \\
0 & 0 & 0 & e^{i\theta} \\
\end{bmatrix}.
\end{equation}

Defining $\Lambda_{~e}^{a}$= $^k\Lambda^a_{~b}\,^{\ell}\Lambda^b_{~c}\,^A\Lambda^c_{~d}\,^{\theta}\Lambda^d_{~e}$,
we have that
\begin{equation}
\Lambda_{~c}^{b}= \begin{bmatrix}
A & A^{-1}E\bar{E} & \bar{E}e^{i\theta} & Ee^{-i\theta}  \\
AB\bar{B} & A^{-1}(1+B\bar{B}E\bar{E}+\bar{B}E+B\bar{E}) & (B\bar{B}\bar{E}+\bar{B})e^{i\theta} & (B\bar{B}E+B)e^{-i\theta}   \\
AB & A^{-1}(BE\bar{E}+E) & (1+B\bar{E})e^{i\theta} & BEe^{-i\theta}  \\
A\bar{B} & A^{-1}(\bar{B}E\bar{E}+\bar{E}) & \bar{B}\bar{E}e^{i\theta} & (1+\bar{B}E)e^{-i\theta} \\
\end{bmatrix} \label{gen_Lorentz1}
\end{equation}
with inverse transformation $\Lambda^{~b}_{c}\equiv(\Lambda^{-1})^{b}_{~c}$ as
\begin{equation}
(\Lambda^{-1})^{b}_{~c}= \begin{bmatrix}
A^{-1}(1+B\bar{B}E\bar{E}+\bar{B}E+B\bar{E}) & A^{-1}E\bar{E} & A^{-1}(-\bar{B}E\bar{E}-\bar{E}) &A^{-1}(-BE\bar{E}-E)  \\
AB\bar{B} & A & -A\bar{B} & -AB   \\
(-B\bar{B}E-B)e^{-i\theta} & -Ee^{-i\theta} & (1+\bar{B}E)e^{-i\theta} & BEe^{-i\theta}  \\
(-B\bar{B}\bar{E}-\bar{B})e^{i\theta} & -\bar{E}e^{i\theta} & \bar{B}\bar{E}e^{i\theta} & (1+B\bar{E})e^{i\theta} \\
\end{bmatrix}\label{gen_Lorentz2}
\end{equation}
A general Lorentz transformations has 6 degrees of freedom, and a repeated Lorentz transformation simply serves to redefine
the parameter functions $\{A, \theta, B, E\}$.

\newpage
\subsection{A brief review of teleparallel gravity}\label{sec:BasicTEGR}

Let us consider TEGR and $f(T)$ theories.
We shall follow the covariant approach and choose a complex null frame.
We shall assume that M is a 4D differentiable manifold M
with coordinates $x^\mu$. Therefore, there exists a non-degenerate coframe
field $\bh^a$ defined on a subset $ U$ of $M$. 
We also assume the existence of a symmetric metric field on M, with coordinates $g_{ab}$
with respect to the coframe field, in order to define  a notion of lengths and
angles.
In addition, we assume the existence of a linear affine connection one form $\bomega^{a}_{\phantom{a}b}$
in order to define covariant differentiation.

The final assumption that the spin-connection is metric compatible implies that the spin-connection is anti-symmetric, $\bomega_{(ab)}=0$, which can be implemented easily without loss of generality
and will automatically satisfied hereafter. The fundamental variables remaining are the 16 elements of the coframe $\bh^a_{\phantom{a}}$ and the 24 elements of the anti-symmetric spin-connection $\bomega^a_{\phantom{a}b}$.  The torsion  associated with the coframe and spin-connection is:
\begin{equation}
T^a_{\phantom{a}\mu\nu}=\partial_\mu h^a_{\phantom{a}\nu}-\partial_\nu h^a_{\phantom{a}\mu}+\omega^a_{\phantom{a}b\mu}h^b_{\phantom{a}\nu}-\omega^a_{\phantom{a}b\nu}h^b_{\phantom{a}\mu}.
\end{equation}

To derive the field equations for teleparallel gravity, we consider a Lagrangian for $f(T)$ teleparallel gravity where the scalar quantity, $T$, is defined as
\begin{equation}
T =
\frac{1}{4} \; T^\rho_{\phantom{\rho}\mu\nu} \, T_\rho^{\phantom{\rho}\mu \nu} +
\frac{1}{2} \; T^\rho_{\phantom{\rho}\mu\nu} \, T^{\nu \mu}_{\phantom{\rho\rho}\rho} -
            T^\rho_{\phantom{\rho}\mu\rho} \, T^{\nu \mu}_{\phantom{\rho\rho}\nu}.
\label{TeleLagra}
\end{equation}
\noindent The scalar $T$ can be written more compactly using the super-potential, $S^a_{~\mu \nu}$, expressed in terms of the torsion tensor
\begin{equation}
S_a^{\phantom{a}\mu\nu}=\frac{1}{2}\left(T_a^{\phantom{a}\mu\nu}+T^{\nu\mu}_{\phantom{\nu\mu}a}-T^{\mu\nu}_{\phantom{\mu\nu}a}\right)-h_a^{\phantom{a}\nu}T^{\rho \mu}_{~~\rho} + h_a^{\phantom{a}\mu}T^{\rho \nu}_{~~\rho},
\end{equation}
\noindent  so that:
\begin{equation}
T=\frac{1}{2}T^a_{\phantom{a}\mu\nu}S_a^{\phantom{a}\mu\nu}.
\end{equation}

The action of the $f(T)$ theory has
\begin{equation}
L_{Grav}(h^a_{\phantom{a}\mu},\omega^a_{\phantom{a}b\mu})=\frac{h}{2\kappa}f(T),
\end{equation}
\noindent where $h$ is the determinant of the vielbein matrix $h^a_{~\mu}$ \cite{Krssak_Saridakis2015}, and a matter part with canonical energy-momentum
$\Theta_a^{~\mu}$, the variation of which, with respect to  $h^a_{\phantom{a}\mu}$, yields:
\begin{eqnarray}
\kappa \Theta_a^{\phantom{a}\mu}&=&
        h^{-1}f_T(T)\partial_v\left(hS_a^{\phantom{a}\mu\nu}\right)+f_{TT}(T)S_a^{\phantom{a}\mu\nu} \partial_v T \nonumber\\
        &&+\frac{1}{2}f(T)h_a^{\phantom{a}\mu} -f_T(T)T^b_{\phantom{a}a\nu}S_b^{\phantom{a}\mu\nu}-f_T(T)\omega^b_{\phantom{a}a\nu}S_b^{\phantom{a}\mu\nu}.\label{EQ2b}
 \end{eqnarray}
The resulting field equation \eqref{EQ2b}, which is {\it Lorentz covariant},
yields the field equations for determining the coframe.

The condition that the curvature vanishes imposes a second differential constraint on the spin-connection and implies that the most general solution to this equation is of the form: \begin{equation}
\omega^a_{\phantom{a}b\mu} = \Lambda^a_{\phantom{a}c}\partial_\mu\Lambda_{b}^{\phantom{a}{c}}\label{solution_omega}
\end{equation}
where $\Lambda^a_{\phantom{a}b}$ is some yet undetermined Lorentz transformation.

There is another constraint on $\bomega^a_{~b}$ arising from \eqref{EQ2b}, in addition to determining the coframe,
from assuming that the canonical energy momentum is symmetric, so that
 $$\Theta_{[ab]} = 0.$$
From the anti-symmetric part of the  field equations  \eqref{EQ2b}, this condition then implies that

\begin{equation}
             0      = f_{TT}(T)S_{[ab]}^{\phantom{[ab]}\nu} \partial_v T.\label{temp2}
\end{equation}
In the case of TEGR, where $f(T) = T$, equation (\ref{temp2}) vanishes. It also vanishes for $T = T_0$, a constant.
For $f(T)_{} \neq T$, it can be shown that the variation of the gravitational Lagrangian by the flat spin-connection is equivalent to the anti-symmetric part of the field equations \eqref{temp2}  \cite{Hohmann:2017duq,Golovnov1}. Thus if the field equations for the spin-connection are satisfied then the field equations for the coframe are guaranteed to be symmetric \cite{Hohmann:2018rwf}.

{The covariant form of the $f(T)$ field equations can be restated purely in a frame basis:
\beq D_c (f_{T} S_a^{~bc}) + f_{T}[S_a^{~bc} T^d_{~cd} + \frac12 S_a^{~cd} T^b_{~cd} + S_d^{~bc} T^d_{~ca}] + \frac12 \delta_a^{~b} f(T) =  \kappa  \Theta_a^{~b} \eeq



\newpage

\section{Symmetry Algorithm}

We shall assume the existence of a single affine symmetry, ${\bf X}$,  which is not an isotropy, satisfying equations (\ref{Intro:FS2}).
It immediately follows that ${\bf X}$ is a Killing vector.
The algorithm can be trivially generalized to include the existence of multiple affine symmetries if there are no isotropies.

\subsection{Algorithm}

\begin{enumerate}

\item Choose coordinates such that any Killing vector(s) ${\bf X}$ and the metric are put into some canonical form (that is, use any isometry present to choose coordinates adapted to the symmetry). For example, if the Lie group of Killing Vectors is one dimensional (not an isotropy) then
\begin{equation}
{{\bf X}} \equiv \frac{\partial}{\partial \chi}, \Rightarrow  g_{\mu\nu} =  g_{\mu\nu}(x^\gamma),
\end{equation}
where the coordinates are $\chi$ and $x^\gamma$ (and where $\gamma = 1,2,3$; that is, the metric does not depend on $\chi$). Henceforth, the coordinates are kept fixed.

If the dimension of the Lie group of affine frame symmetries is greater than one (e.g., spatially homogeneous or spherically symmetric), then once again choose coordinates adapted to the symmetries to express the metric in canonical form.  Additional consideration must be taken when the Lie group of affine frame symmetries has an isotropy (see \cite{MCH} for details).

\item Construct a canonical coframe, ${\tilde{h}}^a$, such that
\begin{equation}
g_{\mu\nu}=\eta_{ab}  {\tilde{h}}^{a}_{\phantom{a}\mu} {\tilde{h}}^{b}_{\phantom{a}\nu}, \quad\mbox{and}\quad \mathcal{L}_{{\bf X}}({\tilde{h}}^a) =0.
\end{equation}

\item Since ${{\bf X}}$ is an affine frame symmetry, there exists a frame, and hence general Lorentz transformations
${\tilde{\Lambda}}^a_{\phantom{a}b}$ and ${\bar{\Lambda}}^a_{\phantom{a}b}$, such that
\begin{equation}
{h}^a_{\phantom{a}\mu} ={\tilde{\Lambda}}^a_{\phantom{a}b}\tilde{h}^b_{\phantom{a}\mu} \label{frame}
\end{equation}
and
\begin{equation}
\omega^a_{\phantom{a}b\mu}={\bar{\Lambda}}^a_{\phantom{a}c} \partial_\mu ({\bar{\Lambda}}_b^{\phantom{a}c})
\label{spin} 
\end{equation}
so that equations (\ref{Intro:FS2}) imply that
\begin{equation}
\mathcal{L}_{{\bf X}}({\tilde{\Lambda}}^a_{\phantom{a}b}) = 0, \label{one} 
\end{equation}
and
\begin{equation}
\mathcal{L}_{{\bf X}}({\bar{\Lambda}}^a_{\phantom{a}c} \partial_\mu ({\bar{\Lambda}}_b^{\phantom{a}c})) = 0. \label{two} 
\end{equation}
\noindent
Since the Lorentz transformations of the form (\ref{one}) imply (\ref{two}), the latter subsume the former.

\item We can now affect another Lorentz transformation to simplify. We could apply a Lorentz transformation
${\tilde{\Lambda}}^a_{\phantom{a}b}$ to set  the frame as the canonical frame ${\tilde{h}}^a$ (and the spin connection is then of the general form (\ref{spin})). Or, we can apply a Lorentz transformation ${\bar{\Lambda}}^a_{\phantom{a}c}$ to set the
spin connection to zero (i.e., choose a proper frame), in which case the frame is of the general form
\begin{equation}
{\bar{{h}}^a_{\phantom{a}\mu} ={\bar{\Lambda}}^a_{\phantom{a}b}\tilde{h}^b_{\phantom{a}\mu}. } \label{hbar}
\end{equation}

For example, if the Lie group of Killing vectors is one dimensional (not an isotropy) then ${\bar{\Lambda}}^a_{\phantom{a}b}$ satisfies
\begin{equation}
 \frac{\partial}{\partial \chi}{\bar{\Lambda}}^a_{\phantom{a}c} \partial_\mu ({\bar{\Lambda}}_b^{\phantom{a}c}) = 0. \label{LTbar} 
\end{equation}

\end{enumerate}

\newpage

\subsection{Spin connection}

With our choice of a complex null frame $ \{ \bh {^a} \} = \{ \bk, \bell, \bm, \bar{\bm} \}$ and our representation for the most general Lorentz transformations (\ref{gen_Lorentz1}) and (\ref{gen_Lorentz2}) with this gauge choice we have $\omega^a_{~b\mu} \equiv \Lambda^{a}_{~c}\,\partial_\mu(\Lambda^{-1})^c_{~b}$ equals to:
\begin{equation}
\begin{bmatrix}
A^{-1}A_{\mu}-B_\mu\bar{E}-E\bar{B}_\mu & 0 & A^{-1}e^{i\theta}(\bar{E}_\mu-\bar{E}B_\mu\bar{E}) &
A^{-1}e^{-i\theta}(E_\mu - E\bar{B}_\mu E)  \\
0 & -A^{-1}A_{\mu}+E\bar{B}_\mu+\bar{E}B_\mu & A\bar{B}_\mu e^{i\theta} & AB_\mu e^{-i\theta}   \\
AB_\mu e^{-i\theta} & A^{-1}e^{-i\theta}(E_\mu - E\bar{B}_\mu E)  & \bar{E}B_\mu-E\bar{B}_\mu+i\theta_\mu & 0  \\
 A\bar{B}_\mu e^{i\theta} &  A^{-1}e^{i\theta}(\bar{E}_\mu-\bar{E}B_\mu\bar{E}) & 0 & -(\bar{E}B_\mu-E\bar{B}_\mu+i\theta_\mu) \\
\end{bmatrix}
\end{equation}
which equals $\omega^a_{~b\mu}(x^{\gamma})$,
since  $ \frac{d}{d\chi}\left
(\Lambda^{a}_{~c}\,\partial_\mu(\Lambda^{-1})^c_{~b}\right )=0$. We note that $\bomega_{(ab)}=0$ is satisfied automatically.
The spin connection is defined in terms of the six arbitrary functions (Lorentz degrees of freedom) $A,\theta,E$ and $B$. 

An alternative expression for the spin connection is:
\begin{equation}
\omega^a_{~b\mu}
=\begin{bmatrix}
Re(\Theta_\mu)& 0 & \Psi^{I}_\mu & \bar{\Psi}^{I}_\mu  \\
0 & -Re(\Theta_\mu) & \bar{\Psi}^{II}_\mu  & \Psi^{II}_{\mu}  \\
\Psi^{II}_{\mu} & \bar{\Psi}^{I}_\mu  & -i Im(\Theta_\mu) & 0  \\
\bar{\Psi}^{II}_\mu & \Psi^{I}_\mu   & 0 & i Im(\Theta_\mu) \\
\end{bmatrix}
\end{equation}
where the (related) complex valued functions are defined by the Lorentz parameter functions:
\begin{eqnarray} 
\Theta_\mu(x^\gamma) &\equiv&  A^{-1}A_{\mu} -  i \theta_{\mu} - 2\bar{E}B_\mu\nonumber\\ 
\Psi^{I}_{\mu}(x^\gamma) &\equiv& A^{-1}e^{i\theta}(\bar{E}_\mu-\bar{E}B_\mu\bar{E})\label{Theta_Psi}\\  
\Psi^{II}_{\mu}(x^\gamma) &\equiv& AB_\mu e^{-i\theta} \nonumber
\end{eqnarray}

\newpage

\section{Solution for the spin connection}

Denoting eqn. [a,b] as the equation that follows from this set of partial differential equations (PDEs) from row a and column b,
for general right-hand sides that are functions of $x^{\gamma}$ alone,
we have from eqns. [2,0], [0,0] and [2,2] the  following equations:\\
$    AB_\mu e^{-i\theta} = G_\mu(x^\gamma)$\\
$ A^{-1}A_{\mu}-B_\mu\bar{E}-E\bar{B}_\mu = H_\mu(x^\gamma)$\\
$ \bar{E}B_\mu-E\bar{B}_\mu+i\theta_\mu = iK_\mu(x^\gamma)$\\
where $K_\mu$ and $H_\mu$ are real and $G_\mu$  is complex
(we recall that  $A$, $\theta$ are real and $B$, $E$ are complex). We shall assume the general case in which $B_{\mu} \neq 0$.

The real part of  eqn. [2,0] can be written as (where $B_\mu = B^r_\mu + iB^i_\mu$ etc.):
$$   B^r_\mu  = A^{-1} (G^r_\mu(x^\gamma)cos{\theta}-G^i_\mu(x^\gamma)sin{\theta}).$$
These 4 linear PDEs for $B^r_\mu$ can then be combined (by cancelling $A$, eliminating $sin(\theta), cos(\theta)$ and considering all degenerate cases) into the form of the linear PDE:
$$   B^r_{~,\chi}  =  F^\gamma({x^\delta}) B^r_{~,{\gamma}}$$
(summing over $\gamma$).
The general solution of this eqn. is of the form $B^r = F(\chi - \mathcal{F}(x^{\gamma}))$ (and similarly for $B^i$). Substituting this back into the
linear PDEs (for $x^{\gamma}$) and eliminating the trigonometric functions
and solving for $sin^2(\theta) + cos^2(\theta) = 1$, then implies that
\be
A(\chi,x^{\gamma}) F^{\prime}(\chi  - \mathcal{F}(x^{\gamma})) = f(x^{\gamma}), \label{prime}
\ee
where a prime denotes an ordinary derivative, and that $\theta_{,\chi}=0$; i.e.,
$\theta=\theta(x^{\gamma})$. Using the $\chi$-component of  eqn. [2,2] with $\theta_\chi = 0$ then yields
$$ E = \frac{\tilde{K}(x^{\gamma})}{F^{\prime}(\chi  - \mathcal{F}(x^{\gamma}))},$$ whence the eqn. [0,0]
yields
$ A^{-1}A_{\mu} = L_{\mu}(x^{\gamma})$.
Integrating for $\mu = \chi$ we obtain $A=\alpha(x^{\gamma})e^{L(x^{\gamma})\chi}$, and subsequently
integrating with $\mu = \gamma$, we find that $L_{,\gamma} = 0$ and so $L = L_0$, a constant.
Eqn. (\ref{prime}) then yields
$$F^{\prime}(\chi  - \mathcal{F}(x^{\gamma})) = \frac{f(x^{\gamma})}{\alpha(x^{\gamma})} e^{- L_0 \chi};   $$
that is, $F^{\prime}$ is separable, and on integration we obtain
$$F(\chi  - \mathcal{F}(x^{\gamma})) = J(x^{\gamma}) e^{- L_0 \chi}.$$
The remaining eqns. are then all automatically satisfied and simply serve to restrict the form of
the functions ${\omega}^a_{~~b\mu}(x^{\gamma})$.

Summarizing and redefining, we have that:\\
$A=\alpha(x^{\gamma})e^{L_0\chi}$, $L_0$ is constant.\\
$B=\beta(x^{\gamma})e^{-L_0\chi}$, $\beta$ is complex\\
$E=\eta(x^{\gamma})e^{L_0\chi}$, $\eta$ is complex\\
$\theta=\theta(x^{\gamma})$ \\
Then we have for $\mu=\chi$:

$\omega^a_{~~b\chi}=\begin{bmatrix}
L_0(1+\beta\bar{\eta}+\eta\bar{\beta}) & 0 &
L_0\alpha^{-1}e^{i\theta}(\bar{\eta}+\bar{\eta}\beta\bar{\eta}) &
L_0 \alpha^{-1}e^{-i\theta}(\eta+\eta
\bar{\beta}\eta)   \\
0 & -L_0(1 + \eta\bar{\beta} + \bar{\eta}\beta) & -L_0\alpha\bar{\beta}e^{i\theta} & -L_0\alpha\beta e^{-i\theta}   \\
-L_0\alpha\beta e^{-i\theta} & L_0 \alpha^{-1}e^{-i\theta}(\eta+\eta
\bar{\beta}\eta)  & -L_0(\bar{\eta}\beta-\eta\bar{\beta}) & 0  \\
-L_0\alpha\bar{\beta}e^{i\theta} & L_0\alpha^{-1}e^{i\theta}(\bar{\eta}+\bar{\eta}\beta\bar{\eta}) & 0 & -L_0(\eta\bar{\beta}-\bar{\eta}\beta) \\
\end{bmatrix}$\\
and for $\mu=\gamma$:\\
$\omega^a_{~~b\gamma}=\footnotesize{\begin{bmatrix}
\alpha^{-1}\alpha_\gamma-\beta_\gamma\bar{\eta}-\eta\bar{\beta_\gamma} & 0
&\alpha^{-1}
 e^{i\theta}(\bar{\eta_\gamma}-\bar{\eta}^2\beta_\gamma)
 & \alpha^{-1} e^{-i\theta}(\eta_\gamma-\eta^2\bar{\beta_\gamma})  \\
0 & -\alpha^{-1}\alpha_\gamma+\eta\bar{\beta_\gamma}+\bar{\eta}\beta_\gamma
& \alpha\bar{\beta_\gamma}e^{i\theta} & \alpha\beta_\gamma e^{-i\theta}   \\
\alpha\beta_\gamma e^{-i\theta} & \alpha^{-1} e^{-i\theta}(\eta_\gamma-\eta^2\bar{\beta_\gamma})
 & \bar{\eta}\beta_\gamma-\eta\bar{\beta_\gamma}+i\theta_\gamma & 0  \\
\alpha\bar{\beta_\gamma}e^{i\theta} & \alpha^{-1} e^{i\theta}(\bar{\eta_\gamma}-\bar{\eta}^2\beta_\gamma)& 0 &\eta\bar{\beta_\gamma}- \bar{\eta}\beta_\gamma-i\theta_\gamma \\
\end{bmatrix}}$\\
We also note that
\\ \\ $\Lambda_{~c}^{b}= \begin{bmatrix}
\alpha e^{L_0\chi} & \alpha^{-1}\eta\bar{\eta} e^{L_0\chi} & \bar{\eta}e^{i\theta} e^{L_0\chi} &
\eta e^{-i\theta} e^{L_0\chi}\\
\alpha\beta\bar{\beta} e^{-L_0\chi} & \alpha^{-1}(1+\beta\bar{\beta}\eta\bar{\eta}+\bar{\beta}\eta+\beta\bar{\eta}) e^{-L_0\chi} & (\beta\bar{\beta}\bar{\eta}+\bar{\beta})e^{i\theta} e^{-L_0\chi} & (\beta\bar{\beta}\eta+\beta)e^{-i\theta} e^{-L_0\chi}  \\
\alpha\beta & \alpha^{-1}(\beta\eta\bar{\eta}+\eta) & (1+\beta\bar{\eta})e^{i\theta} & \beta\eta e^{-i\theta}  \\
\alpha\bar{\beta} & \alpha^{-1}(\bar{\beta}\eta\bar{\eta}+\bar{\eta}) & \bar{\beta}\bar{\eta}e^{i\theta} & (1+\bar{\beta}\eta)e^{-i\theta} \\
\end{bmatrix}$\\\\
in terms of 6 arbritary functions of $x^{\gamma}$ (Lorentz degrees of freedom). Hence this represents {\em a}
general solution.}

\newpage

\subsection{Special solution}
 
The analysis above covers most of the degenerate cases. Let us explicitly consider the degenerate case
$B_{\mu} = 0$ (i.e., $B = B_0$ a constant, corresponding to a global Lorentz transformation, which leaves the spin connection invariant).

The resulting  eqns. [0,0] and [2,2] can be integrated to yield
\begin{equation}
A = \alpha(x^\gamma)  e{^ {L_0 \chi}},~~ \theta = J_0 \chi +j(x^\gamma),  \label{special}
\end{equation}
where $J_0$ is a constant (and $J_0 \neq 0$ otherwise $\theta=\theta(x^{\gamma})$, and this is a subcase of the general case above). The eqn. [0,3], $A^{-1}e^{-i\theta}{E}_\mu = I_{\mu}(x^\gamma)$, once the above expressions for
$A$ and $\theta$ have been substituted in, can then be integrated. Since $J_0 \neq 0$, the $\mu = \chi$ eqn. yields
\begin{equation}
E = e^{(L_0 + i J_0)\chi} \tilde{I}(x^\gamma) +E_0, \label{special2}
\end{equation}
and the $\mu = \gamma$ eqns. imply that $E_0$ is a constant, which corresponds to a global Lorentz transformation and will be omitted hereafter. We shall see later that the case $\tilde{I}(x^\gamma) \neq 0$
generally leads to a contradiction or to a subcase of the general case.

So taking
$\tilde{I}(x^\gamma) = 0$, this effectively leaves the special case
with $A$ and $\theta$ as given above in (\ref{special}), in which only boosts and spins are allowed. In this very special solution, $E=B=0$, and $Re(\Theta_{\mu}) = \{L_0,{ln(\alpha)},_{\gamma} \}$ and
$Im(\Theta_{\mu}) = \{-J_0, -j,_{\gamma} \}$, so that
\begin{equation}
\omega^a_{~~b\mu} = diag[Re(\Theta_{\mu}), -Re(\Theta_{\mu}), -i Im(\Theta_{\mu}), i Im(\Theta_{\mu})],
\label{diag}
\end{equation}
and, in particular, $\omega^a_{~~b\chi} = diag[L_0, -L_0, i J_0, -i  J_0]$.
We note that this special case is equivalent to the general case above with $\beta=\eta=0$ but with $J_0 \neq 0$ so that $\theta_{,\chi} \neq 0$ (whereas in the general case $\theta_{,\chi} = 0$).

There are no further constraints on the geometry. There will be additional constraints from the field equations.
In particular, unless we are considering TEGR ($f(T)=T$) or $T_{,\gamma}=0$, there will be constraints from the antisymmetric field equations, which we shall investigate next.

\subsection{Important integrability condition}

With the assumption of a single affine frame symmetry, the left-hand sides of (\ref{Theta_Psi}) are independent of $\chi$. From a direct computation, an integrability condition for these differential equations (yielding expressions for $A,\theta,E$ and $B$) is:
\begin{equation}
\left( (\Theta_0)^2 + 4\Psi^{I}_{0}\Psi^{II}_{0}\right)_{,\gamma} =0 \quad \Rightarrow \quad \left( (\Theta_0)^2 + 4\Psi^{I}_{0}\Psi^{II}_{0}\right) = \widetilde{M}^2, \label{integrability}
\end{equation}
where $\widetilde{M}$ is some complex valued constant (the square is only introduced for convenience).
We shall exploit this integrability condition later.

\newpage

\section{Antisymmetric field equations}

We wish to study the effects of having a single affine frame symmetry in an $f(T)$ teleparallel theory of gravity,
and particularly the implications of the antisymmetric part of the field equations. We recall that the
equations of motion for the spin connection, which result from varying the action with respect to the spin connection, is equivalent to the antisymmetric part of the
field equations  which result from varying the action with respect to the coframe \cite{Golovnov1}.  Under Lorentz transformations the antisymmetric part of the field equations transform homogeneously. Namely, if for a given frame-connection pair the antisymmetric part of field equations are satisfied, then the Lorentz transformed frame-connection pair also satisfy the antisymmetric part of the field equations

Let us, for illustration, assume that the single affine symmetry is timelike and choose coordinates so that  $\chi =x^0= t$ (the Killing vector is thus $\frac{\partial}{\partial t}$).
In a neighbourhood, we can always find {\em{local}} (synchronous) coordinates (in which the spatial 3-metric only depends on the local spatial coordinates $x^{\gamma} \equiv (x,y,z)$ and hence can be diagonalized), such that
\begin{equation}
g_{ab} = diag\left[ -1, R^2, P^2, Q^2 \right],
\label{metric} 
\end{equation}
where the metric functions $R, P, Q$ are independent of $t$.  In this case, where the frame metric is given by (\ref{gauge}),
the coframe one-form basis $h^a$ ($a=0,1,2,3$) is:
\begin{equation}
h^a = \left[\begin{array}{c}
\frac{1}{\sqrt{2}}[dt+ R(x,y,z) dx] \\
\frac{1}{\sqrt{2}}[dt- R(x,y,z) dx] \\
\frac{1}{\sqrt{2}}[P(x,y,z) dy+ iQ(x,y,z) dz] \\
\frac{1}{\sqrt{2}}[P(x,y,z) dy- iQ(x,y,z) dz]
\end{array}\right]
\end{equation}
Further, the spin connection one-form is:
\begin{equation} 
\omega^{a}_{~b}=\left[\begin{array}{cccc}
Re(\Theta)_\mu dx^\mu & 0 & \Psi^{I}_{~\mu}dx^\mu & \overline{\Psi}^{I}_{~\mu}dx^\mu \\
0 & -Re(\Theta)_\mu dx^\mu & \overline{\Psi}^{II}_{~~\mu}dx^\mu & \Psi^{II}_{~~\mu}dx^\mu \\
\Psi^{II}_{~~\mu}dx^\mu & \overline{\Psi}^{I}_{~\mu}dx^\mu & -Im(\Theta)_\mu dx^\mu & 0 \\
\overline{\Psi}^{II}_{~~\mu}dx^\mu & \Psi^{I}_{~\mu}dx^\mu & 0 & Im(\Theta)_\mu dx^\mu
\end{array}\right] 
\end{equation}
for some complex valued functions $\Theta(x^\gamma),\Psi^{I}_{~\mu}(x^\gamma),\Psi^{II}_{~~\mu}(x^\gamma)$ given by equation (\ref{Theta_Psi}) in terms of the Lorentz parameters $\theta,A,E$ and $B$.


The antisymmetric part of the field equations are given by (\ref{temp2}). These equations are always satisfied if $f_{TT}(T)=0$ or if the torsion scalar is constant.  If $T = const.$, then the field equations for $f(T)$ teleparallel gravity are equivalent to a rescaled version of TEGR (which looks like GR with a cosmological constant and a rescaled coupling constant) \cite{Krssak:2018ywd}.
Assuming $T \not= const.$, then we have
\begin{equation}
S_{[ab]}^{\phantom{[ab]}\nu} \partial_{\nu} T = 0. \label{eq39}
\end{equation}
Since we have a (timelike) affine frame symmetry, $T$ is independent of $t$, so that $\partial_t T = 0$, and the non-trivial equations are:
\begin{equation}
S_{[ab]}^{\phantom{[ab]}\gamma} \partial_{\gamma} T = 0. \label{eq40}
\end{equation}
We note that ${S_{[a3]}}^{\mu}={\overline{S}_{[a2]}}{}^{\mu}$.  The exact expressions for the non-trivial components of $S_{[ab]}^{\phantom{[ab]}\gamma}$ and the form of the Torsion scalar, $T$, are given by:

\newpage

\begin{eqnarray*}
{S_{[01]}}^{x}&=&-\frac{1}{2}\,{\frac {Re(\Psi^{I}_{~2})}{RP}}+\frac{1}{2}\,{\frac {Im(\Psi^{I}_{~3})}{RQ}}-\frac{1}{2}\,{\frac {Re(\Psi^{II}_{~2})}{RP}}-\frac{1}{2}\,{\frac {Im(\Psi^{II}_{~3})}{RQ}}
\\
{S_{[01]}}^{y}&=&\frac{1}{2}\,{\frac {Re(\Psi^{I}_{~0})}{P}}+\frac{1}{2}\,{\frac {Re(\Psi^{I}_{~1})}{RP}}-\frac{1}{2}\,{\frac {Re(\Psi^{II}_{~0})}{P}}+\frac{1}{2}\,{\frac {Re(\Psi^{II}_{~1})}{RP}}
\\
{S_{[01]}}^{z}&=&-\frac{1}{2}\,{\frac {Im(\Psi^{II}_{~0})}{Q}}-\frac{1}{2}\,{\frac {Im(\Psi^{I}_{~0})}{Q}}-\frac{1}{2}\,{\frac {Im(\Psi^{I}_{~1})}{RQ}}+\frac{1}{2}\,{\frac {Im(\Psi^{II}_{~1})}{RQ}}
\\
{S_{[02]}}^{x}&=&-\frac{1}{4}\,{\frac {Re(\Theta_2)}{RP}}-\frac{1}{2}\,{\frac {Re(\Psi^{II}_{~0})}{R}}+\frac{1}{4}\,{\frac {Im(\Theta_3)}{RQ}}+\frac{1}{4}\,{\frac {Q_{,y}}{RQP}}
\\ && \qquad +i \left(\frac{1}{4}\,{\frac {Im(\Theta_2)}{RP}}+\frac{1}{2}\,{\frac {Im(\Psi^{II}_{~0})}{R}}+\frac{1}{4}\,{\frac {Re(\Theta_3)}{RQ}} -\frac{1}{4}\,{\frac {P_{,z}}{RQP}} \right)
\\
{S_{[02]}}^{y}&=&\frac{1}{4}\,{\frac {Re(\Theta_0)}{P}}+\frac{1}{4}\,{\frac {Re(\Theta_1)}{RP}}+\frac{1}{2}\,{\frac {Im(\Psi^{II}_{~3})}{QP}}-\frac{1}{4}\,{\frac {Q_{,x}}{RQP}}
\\ && \qquad +i \left( \frac{1}{2}\,{\frac {Re(\Psi^{II}_{~3})}{QP}}-\frac{1}{4}\,{\frac {Im(\Theta_0)}{P}}-\frac{1}{4}\,{\frac {Im(\Theta_1)}{RP}} \right)
\\
{S_{[02]}}^{z}&=&-\frac{1}{2}\,{\frac {Im(\Psi^{II}_{~2})}{QP}}-\frac{1}{4}\,{\frac {Im(\Theta_0)}{Q}}-\frac{1}{4}\,{\frac {Im(\Theta_1)}{RQ}}
\\ && \qquad +i \left( -\frac{1}{4}\,{\frac {Re(\Theta_0)}{Q}}-\frac{1}{2}\,{\frac {Re(\Psi^{II}_{~2})}{QP}}-\frac{1}{4}\,{\frac {Re(\Theta_1)}{RQ}}+\frac{1}{4}\,{\frac {P_{,x}}{RQP}} \right)
\\
{S_{[12]}}^{x}&=&\frac{1}{2}\,{\frac {Re(\Psi^{I}_{~0})}{R}}-\frac{1}{4}\,{\frac {Re(\Theta_2)}{RP}}-\frac{1}{4}\,{\frac {Im(\Theta_3)}{RQ}}-\frac{1}{4}\,{\frac {Q_{,y}}{RQP}}
\\ && \qquad +i \left( -\frac{1}{4}\,{\frac {Im(\Theta_2)}{RP}}+\frac{1}{4}\,{\frac {Re(\Theta_3)}{RQ}}+\frac{1}{2}\,{\frac {Im(\Psi^{I}_{~0})}{R}} +\frac{1}{4}\,{\frac {P_{,z}}{RQP}}\right)
\\
{S_{[12]}}^{y}&=&-\frac{1}{4}\,{\frac {Re(\Theta_0)}{P}}-\frac{1}{2}\,{\frac {Im(\Psi^{I}_{~3})}{QP}}+\frac{1}{4}\,{\frac {Re(\Theta_1)}{RP}}+\frac{1}{4}\,{\frac {Q_{,x}}{RQP}}
\\ && \qquad +i \left( -\frac{1}{4}\,{\frac {Im(\Theta_0)}{P}}+\frac{1}{4}\,{\frac {Im(\Theta_1)}{RP}}+\frac{1}{2}\,{\frac {Re(\Psi^{I}_{~3})}{QP}} \right)
\\
{S_{[12]}}^{z}&=&\frac{1}{2}\,{\frac {Im(\Psi^{I}_{~2})}{QP}}-\frac{1}{4}\,{\frac {Im(\Theta_0)}{Q}}+\frac{1}{4}\,{\frac {Im(\Theta_1)}{RQ}}
\\ && \qquad +i \left( -\frac{1}{2}\,{\frac {Re(\Psi^{I}_{~2})}{QP}}+\frac{1}{4}\,{\frac {Re(\Theta_0)}{Q}}-\frac{1}{4}\,{\frac {Re(\Theta_1)}{RQ}}-\frac{1}{4}\,{\frac {P_{,x}}{RQP}} \right)
\\
{S_{[23]}}^{x}&=&i \left( -\frac{1}{2}\,{\frac {Im(\Psi^{I}_{~2})}{RP}}-\frac{1}{2}\,{\frac {Re(\Psi^{I}_{~3})}{RQ}}+\frac{1}{2}\,{\frac {Re(\Psi^{II}_{~3})}{RQ}}-\frac{1}{2}\,{\frac {Im(\Psi^{II}_{~2})}{RP}} \right)
\\
{S_{[23]}}^{y}&=&i \left( \frac{1}{2}\,{\frac {Im(\Psi^{I}_{~0})}{P}}+\frac{1}{2}\,{\frac {Im(\Psi^{II}_{~1})}{RP}}-\frac{1}{2}\,{\frac {Im(\Psi^{II}_{~0})}{P}}+\frac{1}{2}\,{\frac {Im(\Psi^{I}_{~1})}{RP}} +\frac{1}{2}\,{\frac {R_{,z}}{RQP}}\right)
\\
{S_{[23]}}^{z}&=&i \left( \frac{1}{2}\,{\frac {Re(\Psi^{I}_{~0})}{Q}}+\frac{1}{2}\,{\frac {Re(\Psi^{I}_{~1})}{RQ}}+\frac{1}{2}\,{\frac {Re(\Psi^{II}_{~0})}{Q}}-\frac{1}{2}\,{\frac {Re(\Psi^{II}_{~1})}{RQ}}-\frac{1}{2}\,{\frac {R_{,y}}{RQP}} \right)
\\
\end{eqnarray*}

\newpage
\begin{eqnarray}
T&=&
4\,{\frac {Im(\Psi^{II}_{~3})\,Re(\Psi^{I}_{~2})}{QP}}-4\,{\frac {Re(\Psi^{II}_{~2})\,Im(\Psi^{I}_{~3})}{QP}}-4\,{\frac {Re(\Psi^{I}_{~3})\,Im(\Psi^{II}_{~2})}{QP}}
\nonumber\\
&&
+4\,{\frac {Im(\Psi^{I}_{~2})\,Re(\Psi^{II}_{~3})}{QP}}+2\,{\frac {Im(\Psi^{II}_{~1})\,Im(\Theta_2)}{RP}}+2\,{\frac {Im(\Psi^{I}_{~1})\,Re(\Theta_3)}{RQ}}
\nonumber\\
&&
-2\,{\frac {Im(\Theta_1)\,Im(\Psi^{II}_{~2})}{RP}}-2\,{\frac {Re(\Psi^{I}_{~1})\,Re(\Theta_2)}{RP}}+2\,{\frac {Re(\Theta_1)\,Re(\Psi^{I}_{~2})}{RP}}
\nonumber\\
&&
+2\,{\frac {Im(\Theta_1)\,Re(\Psi^{II}_{~3})}{RQ}}-2\,{\frac {Re(\Psi^{II}_{~1})\,Im(\Theta_3)}{RQ}}-2\,{\frac {Re(\Theta_1)\,Im(\Psi^{I}_{~3})}{RQ}}
\nonumber\\
&&
+2\,{\frac {Re(\Theta_1)\,Im(\Psi^{II}_{~3})}{RQ}}+2\,{\frac {Im(\Psi^{I}_{~1})\,Im(\Theta_2)}{RP}}-2\,{\frac {Im(\Psi^{II}_{~1})\,Re(\Theta_3)}{RQ}}
\nonumber\\
&&
-2\,{\frac {Im(\Psi^{I}_{~2})\,Im(\Theta_1)}{RP}}-2\,{\frac {Re(\Psi^{II}_{~1})\,Re(\Theta_2)}{RP}}+2\,{\frac {Re(\Theta_1)\,Re(\Psi^{II}_{~2})}{RP}}
\nonumber\\
&&
-2\,{\frac {Re(\Psi^{I}_{~3})\,Im(\Theta_1)}{RQ}}+2\,{\frac {Re(\Psi^{I}_{~1})\,Im(\Theta_3)}{RQ}}-2\,{\frac {Im(\Psi^{II}_{~0})\,Im(\Theta_2)}{P}}
\\
&&
+2\,{\frac {Im(\Psi^{I}_{~0})\,Im(\Theta_2)}{P}}+2\,{\frac {Im(\Psi^{II}_{~0})\,Re(\Theta_3)}{Q}}+4\,{\frac {Re(\Psi^{I}_{~0})\,Re(\Psi^{II}_{~1})}{R}}
\nonumber\\
&&
-4\,{\frac {Re(\Psi^{I}_{~1})\,Re(\Psi^{II}_{~0})}{R}}-2\,{\frac {Im(\Psi^{I}_{~2})\,Im(\Theta_0)}{P}}+2\,{\frac {Re(\Psi^{II}_{~0})\,Re(\Theta_2)}{P}}
\nonumber\\
&&
-2\,{\frac {Re(\Theta_0)\,Re(\Psi^{II}_{~2})}{P}}-2\,{\frac {Re(\Psi^{I}_{~3})\,Im(\Theta_0)}{Q}}+2\,{\frac {Re(\Psi^I_{~0})\,Im(\Theta_3)}{Q}}
\nonumber\\
&&
-2\,{\frac {Re(\Theta_0)\,Im(\Psi^{II}_{~3})}{Q}}+2\,{\frac {Im(\Psi^{I}_{~0})\,Re(\Theta_3)}{Q}}+2\,{\frac {Im(\Theta_0)\,Im(\Psi^{II}_{~2})}{P}}
\nonumber\\
&&
-2\,{\frac {Re(\Psi^{I}_{~0})\,Re(\Theta_2)}{P}}+2\,{\frac {Re(\Theta_0)\,Re(\Psi^{I}_{~2})}{P}}-2\,{\frac {Im(\Theta_0)\,Re(\Psi^{II}_{~3})}{Q}}
\nonumber\\
&&
+2\,{\frac {Re(\Psi^{II}_{~0})\,Im(\Theta_3)}{Q}}-2\,{\frac {Re(\Theta_0)\,Im(\Psi^{I}_{~3})}{Q}}-4\,{\frac {Im(\Psi^{I}_{~0})\,Im(\Psi^{II}_{~1})}{R}}
\nonumber\\
&&
+4\,{\frac {Im(\Psi^{I}_{~1})\,Im(\Psi^{II}_{~0})}{R}}+2\,{\frac {Im(\Psi^{II}_{~0})\,P_{,z}}{QP}}+2\,{\frac {Re(\Psi^{II}_{~0})\,Q_{,y}}{QP}}
\nonumber\\
&&
+2\,{\frac {Im(\Psi^{II}_{~0})\,R_{,z}}{RQ}}+2\,{\frac {Re(\Psi^{II}_{~0})\,R_{,y}}{RP}}+2\,{\frac {Re(\Theta_0)\,Q_{,x}}{RQ}}
\nonumber\\
&&
+2\,{\frac {Re(\Psi^{I}_{~0})\,Q_{,y}}{QP}}-2\,{\frac {Im(\Psi^{I}_{~0})\,R_{,z}}{RQ}}-2\,{\frac {Im(\Psi^{I}_{~0})\,P_{,z}}{QP}}
\nonumber\\
&&
+2\,{\frac {Re(\Psi^{I}_{~0})\,R_{,y}}{RP}}+2\,{\frac {Re(\Theta_0)\,P_{,x}}{RP}}+2\,{\frac {Im(\Psi^{II}_{~3})\,P_{,x}}{RQP}}
\nonumber\\
&&
+2\,{\frac {Im(\Psi^{I}_{~3})\,P_{,x}}{RQP}}-2\,{\frac {Re(\Psi^{I}_{~2})\,Q_{,x}}{RQP}}-2\,{\frac {Im(\Psi^{II}_{~1})\,P_{,z}}{RQP}}
\nonumber\\
&&
-2\,{\frac {Re(\Psi^{II}_{~1})\,Q_{,y}}{RQP}}+2\,{\frac {Re(\Psi^{II}_{~2})\,Q_{,x}}{RQP}}+2\,{\frac {Re(\Psi^{I}_{~1})\,Q_{,y}}{RQP}}
\nonumber\\
&&
-2\,{\frac {Im(\Theta_3)\,R_{,y}}{RQP}}-2\,{\frac {Im(\Psi^{I}_{~1})\,P_{,z}}{RQP}}+2\,{\frac {Im(\Theta_2)\,R_{,z}}{RQP}}
\nonumber\\
&&
-2\,{\frac {R_{,z}\,P_{,z}}{R{Q}^{2}P}}-2\,{\frac {R_{,y}\,Q_{,y}}{RQ{P}^{2}}}-2\,{\frac {P_{,x}\,Q_{,x}}{{R}^{2}QP}}\nonumber
\end{eqnarray}

\newpage

For completeness we include the ``$t$'' components of $S_{[ab]}^{\phantom{[ab]}\mu}$:
\begin{eqnarray*}
{S_{[01]}}^{t}&=&-\frac{1}{2}\,{\frac {Re(\Psi^{I}_{~2})}{P}}+\frac{1}{2}\,{\frac {Im(\Psi^{I}_{~3})}{Q}}+\frac{1}{2}\,{\frac {Re(\Psi^{II}_{~2})}{P}}+\frac{1}{2}\,{\frac {Im(\Psi^{II}_{~3})}{Q}} -\frac{1}{2}\,{\frac {P_{,x}}{RP}}-\frac{1}{2}\,{\frac {Q_{,x}}{RQ}}
\\
{S_{[02]}}^{t}&=&-\frac{1}{4}\,{\frac {Re(\Theta_2)}{P}}+\frac{1}{2}\,{\frac {Re(\Psi^{II}_{~1})}{R}}+\frac{1}{4}\,{\frac {Im(\Theta_3)}{Q}}+\frac{1}{4}\,{\frac {R_{,y}}{RP}}+\frac{1}{4}\,{\frac {Q_{,y}}{QP}}
\\ && \qquad +i \left( \frac{1}{4}\,{\frac {Im(\Theta_2)}{P}}+\frac{1}{4}\,{\frac {Re(\Theta_3)}{Q}}-\frac{1}{2}\,{\frac {Im(\Psi^{II}_{~1})}{R}} -\frac{1}{4}\,{\frac {R_{,z}}{RQ}} -\frac{1}{4}\,{\frac {P_{,z}}{QP}} \right)
\\
{S_{[12]}}^{t}&=&-\frac{1}{2}\,{\frac {Re(\Psi^{I}_{~1})}{R}}+\frac{1}{4}\,{\frac {Re(\Theta_2)}{P}}+\frac{1}{4}\,{\frac {Im(\Theta_3)}{Q}}+\frac{1}{4}\,{\frac {Q_{,y}}{QP}}+\frac{1}{4}\,{\frac {R_{,y}}{RP}}
\\ && \qquad +i \left( \frac{1}{4}\,{\frac {Im(\Theta_2)}{P}}-\frac{1}{2}\,{\frac {Im(\Psi^{I}_{~1})}{R}}-\frac{1}{4}\,{\frac {Re(\Theta_3)}{Q}} -\frac{1}{4}\,{\frac {R_{,z}}{RQ}} -\frac{1}{4}\,{\frac {P_{,z}}{QP}} \right)
\\
{S_{[23]}}^{t}&=&i \left( -\frac{1}{2}\,{\frac {Im(\Psi^{I}_{~2})}{P}}-\frac{1}{2}\,{\frac {Re(\Psi^{I}_{~3})}{Q}}-\frac{1}{2}\,{\frac {Re(\Psi^{II}_{~3})}{Q}}+\frac{1}{2}\,{\frac {Im(\Psi^{II}_{~2})}{P}} \right)
\end{eqnarray*}


\newpage

\subsection{General case}

In general, eqns. (\ref{eq39}) can be considered as 6 real linear eqns. for the 4 quantities $T_{,\mu}$ \cite{GolovnevGuzman21}. If 4 or more of these linear eqns. are linearly independent, then the only solution is
$T_{,\mu}=0$, and hence $T=const.$ This special solution is extremely "fine tuned".
Therefore, the conditions for the existence of non-trivial solutions consist of a number of degenerate cases in which at most 3 of the above eqns. are linearly independent and the ensuing suite of integrability conditions. This is a daunting task, with many special cases to be investigated (including various cases in which one or more of the $T_{,\gamma}$ are zero). Perhaps the most fruitful case occurs when all of the ${S_{[ab]}}^{\gamma} = 0$; we shall investigate this ({\it the}) {\it degenerate} case below.

However, in the case under study here with an affine symmetry (in which $T_{,t}=0$), these eqns. reduce to
eqns. (\ref{eq40}) which are, in principle, 6 real linear eqns. for the 3 quantities $T_{,\gamma}$. So it is possible that some progress can be made in this case. However, even here progress is difficult. As an illustration let us consider the special case $B_{,\mu}=0$ given by eqns. (\ref{diag}).

Using eqns. (\ref{diag}), eqns. (\ref{eq40}) yield one trivial eqn. and the simple eqn.
\begin{equation}
{R_{,z}}  {T_{,y}} - {R_{,x}}{T_{,z}} = 0, \label{intcon1}
\end{equation} 
which has the general solution
\begin{equation}
{T} = T(x,{R}), \label{intcon2}
\end{equation}
where $T(x,{R})$ is an arbitrary function of $x$ and the metric function $R$. Assuming eqn. (\ref{intcon1}),
the remaining 4 linearly dependent eqns. (\ref{eq40}) gives rise to a suite of integrability conditions.
If we eliminate $T_{,x}$ from these eqns. (assuming that $a_{2} \neq 0,~ a_{3} \neq 0$, where here $a \equiv ln(\alpha)$ and $a_{\gamma} \equiv a_{,\gamma}$, and also where $j_{2} \neq  -\frac{P_{z}}{Q},~ j_{3} \neq \frac{Q_{y}}{P}$; any of these conditions allow for the full integration of the eqns. (\ref{eq40}) leading to severe constraints on the geometry), linear combinations of these eqns. using (\ref{intcon1}) yield
\begin{equation}
[a_2 R_{,y} + a_3 \frac{P^2}{Q^2}  R_{,z}] J_0 R + [a_3 R_{,y} - a_2 R_{,z}] a_1 \frac{P}{Q} = 0,
\end{equation} 
and 
\begin{equation}
[-\frac{Q_{x}}{Q}(j_{2}  + \frac{P_{z}}{Q}) -  j_1 \frac{P}{Q}
(j_{3} - \frac{Q_{y}}{P})] R_{,y} + \frac{P}{Q}[j_1\frac{1}{R} (j_{2}  + \frac{P_{z}}{Q}) +
(R L_0 - \frac{P_{x}}{P}) (j_{3} - \frac{Q_{y}}{P})]R_{,z} = 0,
\end{equation} 
which constitute differential eqns. for $a$ and $j$, respectively. The remaining two eqns. yield an additional
differential eqn. for $a, j$ and a partial differential eqn. for   $T(x,{R})$, where in this case
\begin{equation}
(RPQ)~ T(x,{R})=\frac{L_0}{PQ}(\frac{P_{,x}}{P}+\frac{Q_{,x}}{Q}) +(j_3 R_{,y} - j_2 R_{,z}) - [     
\frac{1}{Q} R_{,z} P_{,z} + \frac{1}{P} R_{,y} Q_{,y} + \frac{1}{R} P_{,x} Q_{,x}]. 
\end{equation}

\newpage

\section{Solution of the degenerate antisymmetric field equations}

Solving the linear system of equations in the degenerate case, ${S_{[ab]}}^{\gamma} = 0$,   yields the following solution in terms of the frame functions $R,P,Q$ and some of their derivatives $ R_y, R_z, Q_x, Q_y, P_x, P_z$:
\begin{eqnarray}
Re(\Psi^{II}_{~1})&=&\frac{1}{2}\,{\frac {2\,Re(\Psi^{II}_{~0})\,PR-R_{,y}}{P}},
\\
Im(\Psi^{II}_{~1})&=&\frac{1}{2}\,{\frac {2\,Im(\Psi^{II}_{~0})\,QR-R_{,z}}{Q}},
\\
Re(\Psi^{II}_{~2})&=&-\frac{1}{2}\,{\frac {Re(\Theta_0)\,PR-P_{,x}}{R}},
\\
Im(\Psi^{II}_{~2})&=&-\frac{1}{2}\,Im(\Theta_0)\,P,
\\
Re(\Psi^{II}_{~3})&=&\frac{1}{2}\,Im(\Theta_0)\,Q,
\\
Im(\Psi^{II}_{~3})&=&-\frac{1}{2}\,{\frac {Re(\Theta_0)\,QR-Q_{,x}}{R}},
\\
Re(\Psi^{I}_{~1})&=&-\frac{1}{2}\,{\frac {2\,Re(\Psi^{I}_{~0})\,PR-R_{,y}}{P}},
\\
Im(\Psi^{I}_{~1})&=&-\frac{1}{2}\,{\frac {2\,Im(\Psi^{I}_{~0})\,QR+R_{,z}}{Q}},
\\
Re(\Psi^{I}_{~2})&=&\frac{1}{2}\,{\frac {Re(\Theta_0)\,PR-P_{,x}}{R}},
\\
Im(\Psi^{I}_{~2})&=&\frac{1}{2}\,Im(\Theta_0)\,P,
\\
Re(\Psi^{I}_{~3})&=&\frac{1}{2}\,Im(\Theta_0)\,Q,
\\
Im(\Psi^{I}_{~3})&=&-\frac{1}{2}\,{\frac {Re(\Theta_0)\,QR-Q_{,x}}{R}},
\\
Re(\Theta_1)&=&0,
\\
Im(\Theta_1)&=&0,
\\
Re(\Theta_2)&=&-Re(\Psi^{II}_{~0})\,P+Re(\Psi^{I}_{~0})\,P,
\\
Im(\Theta_2)&=&-{\frac {Im(\Psi^{II}_{~0})\,PQ-Im(\Psi^{I}_{~0})\,PQ-P_{,z}}{Q}},
\\
Re(\Theta_3)&=&-Im(\Psi^{II}_{~0})\,Q-Im(\Psi^{I}_{~0})\,Q
\\
Im(\Theta_3)&=&{\frac {Re(\Psi^{II}_{~0})\,PQ+Re(\Psi^{I}_{~0})\,PQ-Q_{,y}}{P}},
\end{eqnarray}

Substituting the above solution into the remaining super-potential terms yields
\begin{eqnarray*}
{S_{[01]}}^t &=& - Re(\Theta_0)  \\
{S_{[02]}}^t &=&   {Re(\Psi^{II}_{~0})} -i {Im(\Psi^{II}_{~0})} = \overline{\Psi}^{II}_{~~0}\\
{S_{[12]}}^t &=&   {Re(\Psi^{I}_{~0})} +i {Im(\Psi^{I}_{~0})} = \Psi^{I}_{~0}\\
{S_{[23]}}^t &=& - iIm(\Theta_0)
\end{eqnarray*}

The resulting torsion scalar is then of the form
\begin{eqnarray}
T&=&2\left(Re(\Theta_0)\right)^2 - 2\left(Im(\Theta_0)\right)^2 + 8 {Re(\Psi^{I}_{~0})}{Re(\Psi^{II}_{~0})}- 8 {Im(\Psi^{I}_{~0})}{Im(\Psi^{II}_{~0})}\\ \nonumber
&=& 2Re\left((\Theta_0)^2\right)+8Re\left(\Psi^{I}_{~0}\Psi^{II}_{~0}  \right)\label{new_scalar}\\
&\equiv & 2Re(\widetilde{M}^2), \nonumber
\end{eqnarray}
which we observe must be a constant due to the integrability condition found in equation (\ref{integrability}).  If the spacetime geometry has a single affine symmetry, and if $\partial_\gamma T \not =0$ for $\gamma=x,y,z$, then the anti-symmetric part of the field equations for $f(T)$ teleparallel gravity implies the important result that the torsion scalar must be constant, which is a contradiction. So clearly we need to consider all of the various special cases separately (see below).

\newpage


Let us discuss the degenerate solutions in some more detail.
The 18 non-linear eqns. (44)-(61) are first order partial differential eqns. for  six real functions ($\alpha, \theta, \beta, \eta$) of $(x^{\gamma})$ in terms of the functions
($Re(\Theta_0), Im(\Theta_0), Re(\Psi^{I}_{~0}),$  $Im(\Psi^{I}_{~0}),
Re(\Psi^{II}_{~0}),Im(\Psi^{II}_{~0})$), which also depend on the
constant parameter $L_0$ (and are all zero when $L_0=0$). In the general case in which $L_0 \neq 0$,
$\beta_{,\mu} \neq 0$ and $T_{,\mu} \neq 0$, we can manipulate eqns. (44)-(61) to obtain the linear PDE for $\beta$:

\begin{equation}
(\frac{P_{x}}{P} - \frac{Q_{x}}{Q})[\beta_{,1} + L_0 R \beta] + {R}(\frac{R_{,y}}{P} + \frac{i R_{,z}}{Q})[\frac{1}{P}\beta_{,2}   + \frac{i}{Q} \beta_{,3} ] = 0,
\end{equation}   
in terms of metric functions. Solving this eqn. for $\beta$, the remaining eqns. (44)-(61) can then be solved to obtain the complex functions $\bar{\eta}$ and $e^{\xi} \equiv \alpha e^{-i\theta}$, whence eqns (56)-(61) and (50)-(55) serve to define the partial derivatives of ${\xi}$ and $\bar{\eta}$, respectively, leading to a suite of integrability conditions (from the commutation of partial derivatives). In general, these integrability conditions will lead to constraints on the geometry.

As an illustration, let us consider the $L_0=0$ (otherwise general) case, whence all of the eqns. (44)-(61)
now just depend on the metric functions and their derivatives. The reduced eqns. (56)-(61) immediately yields the PDEs

\begin{equation}
(P^2)_{,x}(Q^2)_{,x} \Xi_{1} + (R^2)_{,y}(Q^2)_{,x}\Xi_{2} +  (R^2)_{,z}(P^2)_{,x}\Xi_{3} = 0,
\end{equation}
where $\Xi_{\gamma} = \alpha_{,\gamma}$ or $\Xi_{\gamma} = \theta_{,\gamma} + [0,\frac{P_z}{Q}, -\frac{Q_y}{P}]$.
More importantly, eqns. (44)-(61) yield

\begin{equation}
(P^2)_{,x} (R^2)_{,zy} - (R^2)_{,z} (P^2)_{,xy} = 2P^2[Q_{,x}P_{,xx}   -  P_{,x}Q_{,xx}],
\end{equation}
and further constraints on $R_{,\gamma}$, which severely constrain the geometry.

There are a number of special cases that should be investigated, particularly
when $T_{,\gamma} = 0$ for one of $\gamma = 1, 2, 3$, whence we lose the
corresponding  $[a,b,\gamma]$  eqns. We shall explicitly consider the special case $\beta=0$ below.

\newpage

\subsection{The special case $B=0$}

Let us consider the special case $B_{,\mu}=0$ (i.e., $\Psi^{II}_{\mu}=0$) in which $\omega^a_{~~b\mu}$  is given by
(\ref{diag}) and $g_{\mu \nu}$ is given by (\ref{gauge}).
In the general case in which $T_{,\gamma} \neq 0$ ($\gamma =1-3$), the [a,b,$\gamma$] eqns., namely, $S_{[ab]}^{\phantom{[ab]}\gamma} = 0$, all yield non-trivial eqns. We recall that none of $\{R,P,Q\}$ can vanish.

Eqns. [0,1,$\gamma$] and  [2,3,$\gamma$] immediately yield
$R_{,y} = R_{,z} = 0$, so that $R=R(x)$, and an x-coordinate transformation (redefinition) can be used to set $R=1$.
Therefore, the resulting geometry is severely restricted.

The remaining eqns. [0,1,$\gamma$] and [2,3,$\gamma$], together with the
real and imaginary parts of eqns. [0,2,$\gamma$] and  [1,2,$\gamma$], then yield
conditions on $\Psi^{I}_{\mu} = A^{-1}e^{i\theta}(\bar{E}_\mu-\bar{E}B_\mu\bar{E})$ and hence on $E_{\mu}$,
with the solution $\tilde{I}(x^\gamma)=0$ in (\ref{special2}) (i.e., $E=0$).

The remaining non-trivial eqns. then yield
\begin{equation}
\frac{P_{x}}{P} =  \frac{Q_{x}}{Q},
\end{equation}
and
\begin{equation}
Im{\Theta_0} = 0, ~~ Im{\Theta_1} = 0.
\end{equation}
The first eqn. implies that $J_0 =0$, which means that this case reduces to a subcase of the general case above.
The expression for the torsion scalar $T=T(x,y,z)$ can be computed.

There are special case when $T_{,\gamma} = 0$ for one of $\gamma = 1, 2, 3$, whence we lose the
corresponding  $[a,b,\gamma]$  eqns. Let us consider the two cases $T_{,x} = 0$
and $T_{,y} = T_{,z} = 0$ below

\subsubsection{The subcase $T_{,x} = 0$}

Let us next consider the special case $T_{,x} = 0$ (so that $T=T(y,z)$) in which $B=E=0$. In this case there are no conditions resulting from the $[a,b,x]$ eqns.
If both $T_{,y} = 0$ and $T_{,z} = 0$, then $T=T_0$, and we shall not pursue this case further. Thus, at least one of $T_{,y}$ or $T_{,z}$
is non-zero, which gives rise to the same eqns., denoted [a,b] (eqns.
[a,b,y] or [a,b,z], both of which are identical).

We again immediately find from eqns. [2,3] that
\begin{equation}
  R_y = R_z = 0,
\end{equation}
and hence $R$ can be set to unity.
The real and imaginary parts of eqns. [0,2] and [1,2] then yield

\begin{equation}
Im{\Theta}_0 =0, ~~ Im{\Theta}_1 =0,
\label{II}
\end{equation}
which implies that $J_0=0$ (so that we are back in as a subcase of the general case) and $j=j(y,z)$.
The remaining eqns. field

\begin{equation}
Re{\Theta}_1 =0, ~~ \alpha=\alpha(y,z),
\label{IIb}
\end{equation}
and
\begin{equation}
\frac{P_{x}}{P} = \frac{Q_{x}}{Q} = Re{\Theta}_0 = L_0. \label{Qeqnb}
\end{equation}

The expression for the torsion scalar becomes
\begin{equation}
T = 2L_0^2,
\end{equation}
a constant.

\subsubsection{The subcase $T_{,y} = T_{,z} = 0$}

When $T_{,x} \neq 0$, but $T_{,y} = T_{,z} = 0$ ($T=T(x)$), only the $[a,b,x]$ eqns. are valid.
The $[0,2,x]$ and  $[1,2,x]$ eqns. yield

\begin{equation}
Re{\Theta}_2 =0, ~~ Re{\Theta}_3 =0; ~~ \alpha = \alpha(x),
\label{III}
\end{equation}
and
\begin{equation}
Im{\Theta}_2 = J_y = \frac{P_z}{Q}, ~~ Im{\Theta}_3 = J_z =\frac{Q_y}{P},
\label{IIIb}
\end{equation}
which leads to an integrability condition on the functions $P, Q$ (from $J_{,yz} = J_{,zy}$).

There is no constraint on $Im{\Theta}_0$ (or $J_0$, so that this is a non-trivial subcase not contained in the general case), but the torsion scalar is given by
\begin{equation}
T = 2(L_0^2 - J_0^2),
\end{equation}
which is a constant.

In conclusion, we can see that the special cases above do not lead to viable solutions.

\newpage


\section{Discussion} \label{sec:discussion}

In teleparallel theories of gravity (such as TEGR and $f(T)$ theories) the (co)frame basis and the spin-connection replace the
metric tensor as the primary geometric object of study. Symmetries in teleparallel geometries are represented by
affine frame symmetries. In this paper we studied teleparallel geometries with a single affine symmetry.

We adopted the locally Lorentz covariant approach and utilized a complex null gauge with a complex null frame.
We introduced an algorithm to study geometries with an affine frame symmetry, which consists of
(i) choosing coordinates adapted to the symmetry, (ii) constructing a canonical frame, in which to (iii) solve equations
(\ref{one}) and (\ref{two}) for the spin connection which determines the Lorentz parameter functions ($A, \theta, B, E$).

We presented the solution in the case of a single affine frame symmetry, in which all of the constraints on the geometry are determined. There will be additional constraints
arising from the field equations. In particular, unless we are considering TEGR ($f(T)=T$) or geometries with $T_{,\gamma}=0$, there will be constraints from the antisymmetric field equations.

We explicitly studied the effects of the antisymmetric part of the field eqns. by assuming for definiteness
a single  timelike affine symmetry in which we can always adopt suitable {\em{local}} coordinates.
In this case the 
antisymmetric field eqns. (\ref{eq40}) reduce to  5 real linear eqns. for the 3 quantities $T_{,\gamma}$. In general,
if 3 or more of these linear eqns. are linearly independent, then the unique solution is
$T_{,\mu}=0$, and hence $T=const.$ 
The conditions for the existence of non-trivial solutions consist of a number of special cases in which at most 2 of the above eqns. are linearly independent, each case of which gives rise to a suite of integrability conditions. 
We studied these conditions and partially integrated them in the special case $B=0$ given by eqns. (\ref{diag}). We also found that
the non-trivial components of the antisymmetric part of the field eqns. in the degenerate case,  ${S_{[ab]}}^{\gamma} = 0$, constitute a linear system of PDEs, whose solution generally gives rise to
severe constraints on the geometry. The geometries with $T=const.$  again play a prominent role. In 
particular, the
special case $B=0$, which was treated separately, was shown to not lead to viable geometries.
Finally, we note that
exact solutions are then obtained by applying all of the fields eqns. (which depend on the form of the source assumed), which leads to further restrictions on the geometry.

In future work we shall study teleparallel geometries with multiple symmetries having no isotropies \cite{HMC} (e.g., simply transitive spatially homogeneous spacetimes, and especially the Abelian case of Bianchi type I) and spacetimes with isotropies \cite{MCH} (e.g., spherically symmetric spacetimes).

\section*{Acknowledgments}
The authors would like to thank D. McNutt for helpful communications.  AAC and RvdH were supported by the Natural Sciences and Engineering research Council of Canada.





\end{document}